\documentclass[amsmath,amssymb,twocolumn,showpacs,aps,prb,showkeys,superscriptaddress, 10pt]{revtex4-2} 

\usepackage{graphicx}
\usepackage{ifthen}
\usepackage{xcolor}
\usepackage{bm}
\usepackage{braket}
\usepackage{multirow}
\usepackage{hyperref}
\usepackage{soul}
\usepackage[normalem]{ulem}
\usepackage{multirow}
\hypersetup{%backref=true,
 pdfnewwindow=true, colorlinks=true,
 linkcolor=blue, anchorcolor=blue,
 citecolor=blue, filecolor=blue,
 menucolor=blue, urlcolor=blue}

\def\ve{\varepsilon}
\def\tc{T_{\rm c}}
\def\ef{\ve_{\rm F}}

\def\o{\omega}
\def\op{\omega^\prime}
\def\a{\alpha}

\begin{document}

\title{Nonadiabatic and anharmonic effects in high-pressure H$_3$S and D$_3$S superconductors}

\author{Shashi B. Mishra}
\email{mshashi125@gmail.com}
\affiliation{Department of Physics, Applied Physics and Astronomy, Binghamton University-SUNY, Binghamton, New York 13902, USA}
\author{Elena R. Margine}
\email{rmargine@binghamton.edu}
\affiliation{Department of Physics, Applied Physics and Astronomy, Binghamton University-SUNY, Binghamton, New York 13902, USA}
\date{\today}

\begin{abstract}
Superconductivity in compressed H$_3$S arises from the interplay between high-frequency phonons and a pronounced van Hove singularity near the Fermi level. Using first-principles calculations, we investigate the superconducting properties of H$_3$S and D$_3$S at 160 and 200~GPa, explicitly incorporating anharmonic lattice dynamics and first-order vertex corrections to electron-phonon (e–ph) interactions, thereby going beyond the Migdal approximation underlying conventional Migdal-Eliashberg theory. We find that both anharmonicity and nonadiabatic vertex corrections suppress the effective e-ph coupling and reduce the superconducting critical temperature ($\tc$). Calculations performed within the energy-dependent full-bandwidth Eliashberg formalism, including both anharmonic and vertex effects, yield $\tc$ values in close agreement with experimental measurements for D$_3$S at both pressures and for H$_3$S at 200~GPa. 
\end{abstract}

\maketitle

\section{\label{sec:intro}Introduction}

The quest for room-temperature superconductivity has accelerated after the discovery of high critical temperatures ($\tc$) in H$_3$S under extreme pressure~\cite{Drozdov2015}. Subsequent studies have revealed several other hydrogen-rich compounds as promising candidates for near-room-temperature superconductivity~\cite{Somayazulu2019,Drozdov2019,Kong2021,Troyan2021,Ma2022}, and more recently, evidence of room-temperature superconductivity has been reported in the La-Sc-H system~\cite{Song2025}. These advances have been enabled by major developments in high-pressure synthesis and characterization techniques, pioneered largely by M. Eremets and collaborators~\cite{Drozdov2019,Kong2021,Einaga2016,Du2025}. In parallel, \textit{ab initio} predictions have guided experimental efforts by identifying candidate structures and elucidating superconducting properties in hydrogen-rich materials~\cite{Duan2014,Wang2012,Errea2015,Liu2017,Peng2017,Pickard2020,Cataldo2021,Hilleke2022,He2024,Lucrezi2024,Renskers2025}. 

The theoretical basis of hydride superconductivity can be traced back to Ashcroft's hypothesis~\cite{Ashcroft1968} that metallic hydrogen could achieve high-$\tc$ through conventional electron-phonon (e-ph) coupling~\cite{Bardeen1957}. Recognizing the experimental challenges posed by the high pressures required to metallize pure hydrogen, Ashcroft subsequently proposed that hydrogen-rich materials, in which hydrogen atoms experience chemical precompression from heavier elements, could become metallic and superconducting under experimentally accessible conditions, broadening the search to hydrogen-containing alloys and compounds~\cite{Ashcroft2004}. A defining feature of these hydrides is the combination of high vibrational frequencies and strong e-ph coupling, which underpins their high-$\tc$ behavior. 

Within the hydride family, H$_3$S has emerged as the prototypical system, in which superconductivity is driven by two key ingredients: the exceptionally high-frequency vibrations of hydrogen atoms (maximum phonon frequency, $\omega_{\rm ph} \approx 200-250$~meV) and a pronounced van Hove singularity (vHs) located near the Fermi level. The narrow width of the vHs leads to a relatively small Fermi velocity, (2.1$-$3.7)$\times10^5$~m/s~\cite{Talantsev2019}, and hence a low Fermi energy scale ($\ef \approx 40-50$~meV~\cite{Gorkov2018}). Together with the large phonon frequencies, this results in a high adiabatic ratio, $\eta = \omega_{\rm ph}/ \ef$, which challenges the validity of Migdal's approximation~\cite{Migdal1958,Jarlborg2016} and necessitates the inclusion of vertex corrections for an accurate theoretical description of the e-ph interaction ~\cite{Sano2016,Mishra2025}. The light mass of hydrogen further amplifies anharmonic lattice vibrations, which not only stabilize the high-pressure structure~\cite{Errea2016} but also strongly renormalize the phonon spectrum and e-ph coupling strength~\cite{Errea2015,Sano2016,Mishra2025}.

The interplay between anharmonic and nonadiabatic effects becomes particularly critical when comparing theoretical predictions with experimental results. Experimental tunneling and resistivity measurements have reported superconducting gaps and transition temperatures in samples near 150$-$161~GPa range~\cite{Du2025}, whereas most theoretical studies have focused on higher pressures (200$-$250~GPa), often neglecting anharmonic contributions. In our recent work at 200~GPa~\cite{Mishra2025}, we demonstrated that accounting for both vertex corrections and quantum anharmonic effects significantly modifies the e-ph coupling and lowers the predicted $\tc$, bringing theoretical predictions into close agreement with experimental observations. These findings emphasize the importance of systematic \textit{ab initio} investigations that incorporate anharmonicity and vertex corrections to fully capture the pressure-dependent superconductivity of H$_3$S and its isotopic counterparts.

Here, we extend our previous work~\cite{Mishra2025} to perform a comparative analysis of H$_3$S and D$_3$S at 160 and 200~GPa, examining how e-ph coupling, vertex corrections, and anharmonic phonon renormalization evolve with pressure. We find that anharmonic effects are essential for maintaining structural stability at 160~GPa, in agreement with a previous study~\cite{Errea2016}. In terms of superconducting properties, both H$_3$S and D$_3$S exhibit a similar pressure dependence, with the e-ph coupling strength $\lambda$ decreasing slightly as pressure increases, while the e-ph vertex correction contribution $\lambda^{\rm V}$ nearly doubles, leading to a reduction in the superconducting transition temperature $\tc$.

\section{\label{sec:results}Results and Discussion}

%\subsection{\label{sec:ph}Electronic and vibrational properties}

\begin{figure}[!t]
    \centering 
    \includegraphics[width=\linewidth]{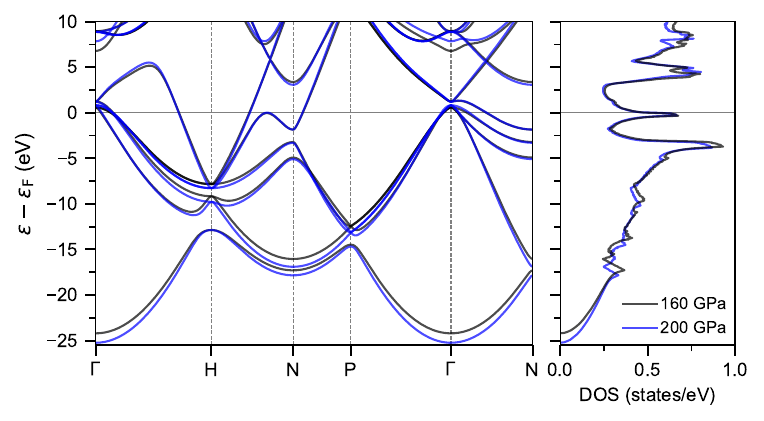}
    \caption{Electronic band structure and density of states (DOS) of H$_3$S at 160~GPa (blue line) and 200~GPa (black line).} 
    \label{fig:dos}
\end{figure}

H$_3$S adopts the body-centered cubic $Im\bar{3}m$ phase under high pressure~\cite{Einaga2016,Goncharov2017}, which remains experimentally stable over a broad pressure range (150$-$300~GPa). The superconducting $\tc$ reaches its maximum near 150$-$160~GPa and gradually decreases with further compression~\cite{Drozdov2015,Minkov2020}. Our \textit{ab initio} calculations yield optimized lattice constants of 3.049~\AA{} at 160~GPa and 2.986~\AA{} at 200~GPa, in good agreement with previous theoretical~\cite{Duan2014,Errea2015} and experimental~\cite{Drozdov2015} reports. The electronic density of states (DOS) at the Fermi level ($\ef$) remains high at both pressures, characterized by a sharp vHs peak near $\ef$ (see Figure~\ref{fig:dos}). With the increase in pressure, the DOS at $\ef$ remains virtually unchanged, while the free-electron-like band edge shifts to lower energies, reflecting a modest broadening of the electronic bandwidth.

\begin{figure*}[!t]
    \centering 
    \includegraphics[width=0.92\linewidth]{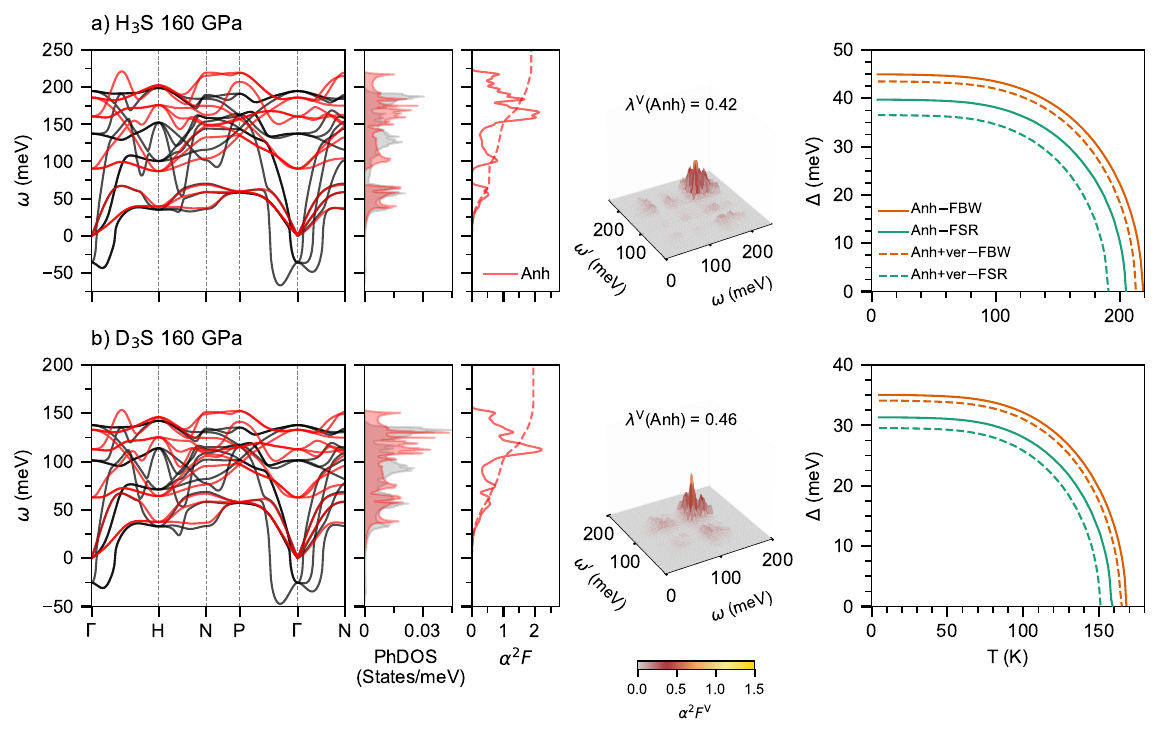}
    \caption{a) Phonon dispersion and phonon density of states (PhDOS) of H$_3$S at 160~GPa calculated within the harmonic (black) and anharmonic (red) approximations. The Eliashberg spectral function $\alpha^2F(\omega)$ (solid lines) and its cumulative e-ph coupling $\lambda(\omega)$ (dashed lines) are shown along with the spectral function from vertex corrections $\a^2F^{\rm V}(\o, \op)$ and its integrated coupling strength $\lambda^{\mathrm{V}}$. The isotropic superconducting gap $\Delta(T)$ is computed using four approaches with anharmonic phonons and an effective Coulomb pseudopotential $\mu_{\mathrm{E}}^* = 0.16$: {\small FBW} (orange), {\small FSR} (teal-green), vertex-corrected {\small FBW} (dashed orange), and vertex-corrected {\small FSR} (dashed teal-green). b) Corresponding results for D$_3$S at 160~GPa. }  
    \label{fig:phP160}
\end{figure*}

At 160~GPa, the harmonic phonon dispersion of H$_3$S shows imaginary phonon modes around the $\Gamma$-point, indicating a saddle point on the Born-Oppenheimer potential energy surface~\cite{Errea2016} (Figure~\ref{fig:phP160}). To account for quantum zero-point motion and anharmonic effects, we employed the anharmonic special displacement method (A-SDM) implemented in the ZG module of EPW~\cite{Zacharias2020,Lee2023}. With anharmonicity included, all phonon modes become real, confirming the dynamical stability of the $Im\bar{3}m$ phase at 160~GPa, consistent with previous work~\cite{Errea2016}. The anharmonic spectrum further reveals a noticeable hardening of the high-frequency H-derived modes, which shift upward by approximately 22~meV. For D$_3$S, the $Im\bar{3}m$ phase is likewise stabilized once anharmonic effects are considered, but the corresponding high-phonon frequencies shift by only about 11~meV due to the larger atomic mass of deuterium. The strongest renormalization occurs for the H-S bond-stretching vibrations, in which hydrogen atoms move toward neighboring sulfur atoms. This pronounced anharmonicity originates from the shallow energy landscape near the second-order quantum phase transition between the $R\bar{3}m$ and $Im\bar{3}m$ phases~\cite{Errea2016}. 

\begin{figure*}[!t]
    \centering 
    \includegraphics[width=\linewidth]{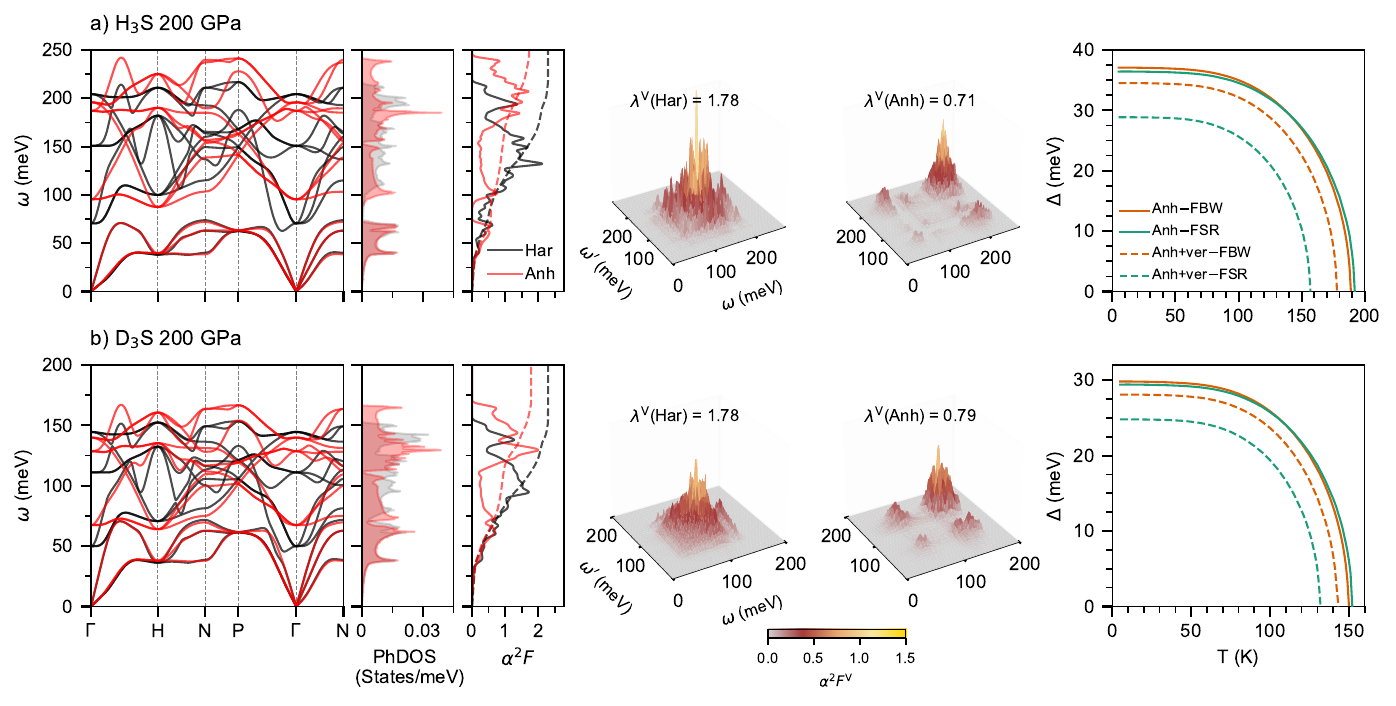}
    \caption{a) Phonon dispersion and phonon density of states (PhDOS) of H$_3$S at 200~GPa calculated within the harmonic (black) and anharmonic (red) approximations. The Eliashberg spectral function $\alpha^2F(\omega)$ (solid lines) and its cumulative e-ph coupling $\lambda(\omega)$ (dashed lines) are shown along with the spectral function from vertex corrections $\a^2F^{\rm V}(\o, \op)$ and its integrated coupling strength $\lambda^{\mathrm{V}}$. The isotropic superconducting gap $\Delta(T)$ is computed using four approaches with anharmonic phonons and an effective Coulomb pseudopotential $\mu_{\mathrm{E}}^* = 0.16$: {\small FBW} (orange), {\small FSR} (teal-green), vertex-corrected {\small FBW} (dashed orange), and vertex-corrected {\small FSR} (dashed teal-green). b) Corresponding results for D$_3$S at 200~GPa.
    }
    \label{fig:P200}
\end{figure*}

At 200~GPa, the harmonic dispersion are positive throughout the Brillouin zone, indicating the enhanced lattice rigidity and dynamical stability of the $Im\bar{3}m$ phase even without anharmonic corrections (Figure~\ref{fig:P200}). In H$_3$S, a small gap separates low-frequency sulfur-derived from high-frequency hydrogen-derived branches around 75~meV, and the acoustic modes display the typical linear dispersion near the $\Gamma$-point. Including anharmonic effects further hardens the H-derived modes by about 25~meV, while D$_3$S shows a smaller shift of roughly 14~meV. Comparison of the 160 and 200~GPa results demonstrates how pressure and isotope substitution tune the phonon spectrum, setting the stage for corresponding changes in the e-ph coupling that ultimately govern the superconducting behavior of H$_3$S and D$_3$S.

To analyze the e-ph interaction, we compute the isotropic Eliashberg spectral function $\alpha^2F(\o)$ and the cumulative coupling strength $\lambda(\o)$ for H$_3$S and D$_3$S (Figures~\ref{fig:phP160} and~\ref{fig:P200}). At 160~GPa, the H$_3$S Eliashberg spectral function displays a small peak in the acoustic region (below 75~meV), contributing about 30\% of the total e-ph coupling, while the remainder comes from the high-frequency H-derived modes, resulting in a total $\lambda$ that saturates at 1.88. The vertex-corrected spectrum $\alpha^2F(\o, \op)$ is also concentrated at high phonon energies above 170~meV, yielding an integrated vertex coupling $\lambda^{\rm V} = 0.42$. D$_3$S at 160~GPa exhibits the same dependence but with isotope-shifted frequencies. The main peak in $\alpha^{2}F(\omega)$ is lowered in energy relative to H$_3$S, while $\lambda(\omega)$ rises to a similar total value, $\lambda=1.94$. Vertex contributions remain concentrated at higher phonon energies and integrate to $\lambda^{\mathrm{V}}= 0.46$.

Although harmonic phonons are dynamically stable at 200~GPa, anharmonicity significantly renormalizes the hydrogen-derived modes, redistributing $\alpha^2F(\omega)$, particularly above 50~meV. As a result, the integrated e-ph coupling $\lambda$ drops by about 25\%, from 2.29 to 1.72~\cite{Errea2015,Mishra2025}. The reduction is even more pronounced for the vertex-corrected spectrum $\alpha^2F(\omega,\omega’)$, where anharmonicity suppresses $\lambda^{\rm V}$ by nearly 60\%, from 1.78 to 0.71. D$_3$S follows the same trend, albeit with slightly smaller shifts  consistent with its heavier isotope mass, with $\lambda$ decreasing from 2.29 to 1.78 and $\lambda^{\rm V}$ from 1.78 to 0.79. Comparing the anharmonic results at the two pressures, $\lambda$ decreases only marginally, while $\lambda^{\rm V}$ nearly doubles as pressure rises from 160 to 200~GPa for both H$_3$S and D$_3$S, as summarized in Table~\ref{tab:supercond}. This increase in $\lambda^{\rm V}$ correlates with a moderate enhancement of the adiabatic parameter $\eta$, which rises from approximately 4.5$-$5.6 at 160~GPa to 5.0$-$6.2 at 200~GPa. The enhancement originates primarily from the anharmonic hardening of $\omega_{\rm ph}$ from about 225 to 250~meV, while the Fermi energy remains nearly constant. Upon isotopic substitution, $\eta$ in D$_3$S also increases slightly, from about $3.0-3.8$ to $3.3-3.8$ over the same pressure range.

\begin{figure}
    \centering
    \includegraphics[width=0.9\linewidth]{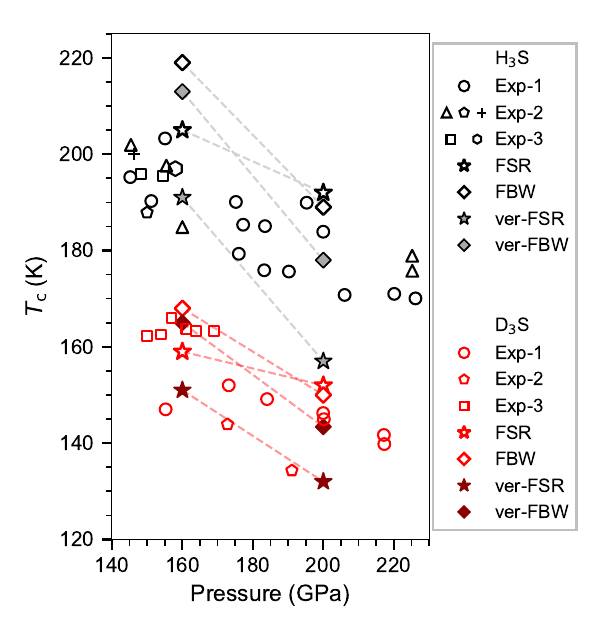}
    \caption{Superconducting critical temperatures ($\tc$) calculated with anharmonic phonons for H$_3$S and D$_3$S at 160 and 200~GPa using four approaches: {\small FSR} (open star), {\small FBW} (open diamond), ver-{\small FSR} (filled star), and ver-{\small FBW} (filled diamond). Experimental data extracted from the literature are shown for comparison and labeled as Exp-1~\cite{Drozdov2015}, Exp-2~\cite{Einaga2016,Mozaffari2019,Nakao2019,Eremets2016}, and Exp-3~\cite{Du2025,Minkov2020}. Black symbols represent H$_3$S and red symbols represent D$_3$S.} 
    \label{fig:tc_comparison}
\end{figure}

Building on the phonon and e-ph coupling analysis, we examine the superconducting properties of H$_3$S and D$_3$S. To this end, we solve the isotropic Eliashberg equations for the superconducting gap $\Delta(T)$ using both the Fermi-surface-restricted (FSR) and full-bandwidth (FBW) approaches~\cite{Lucrezi2024,Mishra2025,Lee2023}. In the FSR approximation, the density of states is treated as constant near $\ef$, whereas the FBW formalism explicitly retains its energy dependence. All calculations include anharmonic phonons and nonadiabatic vertex corrections, and use a Coulomb pseudopotential of $\mu^*_{\mathrm{E}}=0.16$. The isotropic treatment is justified by prior superconductivity calculations using the anisotropic Eliashberg formalism, which showed that H$_3$S remains a single-gap, weakly anisotropic superconductor in the 180$-$200~GPa pressure range~\cite{Lucrezi2024,Flores2016}. This conclusion is further supported by recent tunneling spectroscopy and upper critical field measurements, which indicate that at 158~GPa, H$_3$S exhibits a nearly isotropic superconducting gap characteristic of strong-coupling conventional superconductivity~\cite{Du2025}. 
Figures~\ref{fig:phP160} and \ref{fig:P200} show the temperature dependence of the superconducting gap $\Delta (T)$, where the critical temperature $\tc$ is defined as the point at which $\Delta$ vanishes. A summary of the computed transition temperatures and their comparison with experimental and previous theoretical results is provided in Figure~\ref{fig:tc_comparison} and Table~\ref{tab:supercond}.
For H$_3$S, adiabatic FSR calculations yield $\tc=205$~K, close to the upper limit of the experimental range (184$-$203~K) reported for samples measured between 150$-$161~GPa~\cite{Drozdov2015,Einaga2016,Du2025,Minkov2020,Mozaffari2019,Nakao2019}. At this pressure, the FBW estimate is slightly higher but remains comparable to earlier theoretical predictions based on the FSR approach~\cite{Errea2016}. Including vertex corrections reduces $\tc$, giving 191~K for ver-FSR and 213~K for ver-FBW, with the former lying near the middle of the experimental range. Substituting hydrogen with deuterium lowers the critical temperature by approximately 40-50~K, depending on the computational method employed.
For D$_3$S, the ver-FBW result ($\tc=165$~K) shows excellent agreement with experimental measurements of 163$-$166~K obtained between 150$-$161~GPa in Ref.~\cite{Minkov2020}. The reduction in $\tc$ for D$_3$S is directly linked to the lower characteristic phonon frequency $\omega_{\rm log}$. While the e-ph coupling is only weakly isotope dependent, the characteristic phonon frequency scales approximately with the inverse square root of the atomic mass, decreasing by 25~meV in D$_3$S relative to H$_3$S, as shown in Table~\ref{tab:supercond}. At 200~GPa, the ver-FBW approach yields 178~K for H$_3$S and 143~K for D$_3$S. These values again fall within the corresponding experimental ranges of 175$-$185~K for measurements between 192$-$208~GPa in H$_3$S~\cite{Drozdov2015,Einaga2016,Eremets2016} and 142$-$146~K for measurements between 195$-$205~GPa in D$_3$S~\cite{Drozdov2015}. %

\begin{table*}[!t]
    \caption{Electron-phonon coupling strength ($\lambda$), vertex-corrected e-ph coupling strength ($\lambda^{\rm V}$), logarithmic average phonon frequency ($\omega_{\log}$), and superconducting critical temperature ($\tc$) obtained from the FSR and FBW approaches, with and without vertex corrections for H$_3$S and D$_3$S, incorporating anharmonic phonons at 160~GPa and both harmonic and anharmonic phonons at 200~GPa. All superconductivity calculations employ a Coulomb pseudopotential $\mu_{\rm E}^*$=0.16. Experimental $\tc$ values are taken from measurements in the 150--161~GPa~ \cite{Drozdov2015,Einaga2016,Du2025,Minkov2020,Mozaffari2019,Nakao2019,Eremets2016} and 192--208~GPa pressure ranges for H$_3$S, and 150--165~GPa and 195--205~GPa pressure ranges for D$_3$S~\cite{Drozdov2015,Einaga2016,Minkov2020}. Theoretical FSR results from Refs.~\cite{Errea2015,Errea2016} are included for comparison. All $\tc$ values are given in K.}
\label{tab:supercond}
\setlength\tabcolsep{0pt} % let LaTeX compute intercolumn whitespace
\renewcommand{\arraystretch}{1.2}
\smallskip 
\begin{tabular*}{\textwidth}{@{\extracolsep{\fill}}l c c c c c c c c c c c c}
\hline\hline \noalign{\vskip 1mm}
Element &Pressure & Phonons & $\lambda$ & $\lambda^{\rm V}$ & $\omega_{\rm log}$& \multicolumn{2}{c}{W/o vertex correction} & \multicolumn{2}{c}{With vertex correction} & $T_{\rm c}^{\rm Exp}$ & $T_{\rm c}^{\rm FSR}$ \\
  \cline{7-8} \cline{9-10}
   & (GPa) & &  &  & (meV) & $T_c^{\rm FSR}$ &$T_c^{\rm FBW}$ & $T_c^{\rm ver-FSR}$ & $T_c^{\rm ver-FBW}$ & & ~\cite{Errea2015,Mishra2025}
  \\ \noalign{\vskip 0.5mm}
\hline 
\multirow{3}{*}{H$_3$S} & \multirow{1}{*}{160} & Anh & 1.88 & 0.42 & 107 & 205 & 219 & 191 & 213 & 184$-$203 & 213$-$215\\
\cline{2-12}
& \multirow{2}{*}{200} & Har & 2.29 & 1.78 &107 & 246 & 235 & 198 & 222 & \multirow{2}{*}{175$-$185} & 250 \\ 
 &  & Anh & 1.72 & 0.71 &110 & 192 & 189 & 157 & 178 &  & 194$-$198\\
 \hline
\multirow{3}{*}{D$_3$S} & \multirow{1}{*}{160} & Anh & 1.94 & 0.46 & 82 & 159 & 168 & 151 & 165 & 163$-$166 & 160 \\
\cline{2-12}
& \multirow{2}{*}{200} & Har & 2.29 & 1.78 & 79 & 180 & 174 & 157 & 166 & \multirow{2}{*}{142$-$146} & 183\\ 
 &  & Anh & 1.78 & 0.79 & 85 & 152 & 150 & 132 & 143 & & 152 \\
%    [1mm] \noalign{\vskip 0.05mm}  
\hline\hline
\end{tabular*}
\end{table*}

At 200~GPa, we also compute the superconducting gap using harmonic phonons for both isotopes (see Supplementary Figure~S1 and Table~\ref{tab:supercond}). In this case, the larger e-ph coupling and comparable logarithmic phonon frequency lead to a systematic overestimation of $\tc$ relative to experiment. Even after including vertex corrections, the harmonic FBW and FSR approaches yield values well above the measured ranges. Among all methods tested, the ver-FBW scheme with anharmonic phonons provides the most consistent description for both H$_3$S and D$_3$S, demonstrating that an accurate treatment of phonon anharmonicity, the energy-dependent DOS near Fermi level, and e-ph vertex corrections is essential for a realistic description of superconductivity in hydrogen-rich compounds. 

Several key trends emerge from the data in Table~\ref{tab:supercond}. First, D$_3$S exhibits weaker vertex effects than H$_3$S, reflecting its lower maximum phonon frequency and correspondingly smaller $\alpha^{\mathrm{ver}}$. Second, when both vertex corrections and anharmonic effects are included, the FSR-derived $\tc$ values remain systematically below those from the FBW approach. As discussed in Ref.~\cite{Mishra2025}, this difference arises because, in the ver-FSR formulation, the Eliashberg equations are solved using only the value of the DOS at the Fermi level. In H$_3$S and D$_3$S, the presence of the vHs near $\ef$ enhances this local DOS contribution, whereas the ver-FBW framework captures the full energy dependence of the DOS, yielding a more accurate and physically complete description.

Finally, we examine the isotope coefficient, defined as $\alpha = - \ln(T_c^{\rm D_3S}-T_c^{\rm H_3S}) /{\ln(M^{\rm D_3S}-M^{\rm H_3S})}$, where $M^{\rm D_3S}$ and $M^{\rm H_3S}$ are the corresponding isotopic masses. The dependence of $\alpha$ on the phonon treatment and computational approach used to evaluate the superconducting properties is summarized in Table~S1 of the Supplemental Material~\cite{SI}. At 160~GPa, our calculations yield $\alpha = 0.34$$-$0.38, slightly above the experimental range of 0.22$-$0.34~\cite{Drozdov2015,Minkov2020}. At 200~GPa, $\alpha$ decreases to 0.25$-$0.34, in good agreement with the measured values of 0.23$-$0.38. In contrast, harmonic phonon calculations predict substantially larger $\alpha$ values (0.34$-$0.45). Errea \textit{et al.}~\cite{Errea2015,Errea2016} reported a similar trend, obtaining $\alpha = 0.42$ at 160~GPa and 0.37$-$0.45 for anharmonic and harmonic phonons at 200~GPa. Other studies~\cite{Szczesniak2017,Durajski2016} also found  $\alpha$ values between 0.35 and 0.47, although direct comparison is complicated by differences in the adopted Coulomb pseudopotential $\mu^*$ used to reproduce the experimental $\tc$.

This ratio exceeds the experimentally observed $\Delta(0)$ values of 29$-$31.5~meV for samples measured around 151$-$161~GPa, corresponding to a BCS ratio of
3.55~\cite{Du2025}. Although the experimental value suggests a weak-coupling limit~\cite{Bardeen1957}, H$_3$S is in fact a strong-coupling superconductor. The discrepancy might be attributed to uncertainties or differences in the experimental extraction and fitting procedures for the superconducting gap methods used to interpret tunneling spectra, as well as the limitations of our theoretical calculations. Further theoretical calculations using different exchange-correlation functionals~\cite{Komelj2015}, and systematic experimental studies over a wider pressure range might be useful. 
\section{Conclusion}

We investigated the influence of phonon anharmonicity and first-order e-ph vertex corrections on superconductivity in H$_3$S and D$_3$S at 160 and 200~GPa by solving the Eliashberg equations using both the FSR (constant DOS) and FBW (energy-dependent DOS) formalisms. Anharmonicity serves two key roles: it stabilizes the  $Im\bar{3}m$ phase at 160~GPa and hardens the high-frequency hydrogen-derived modes, which lowers the e-ph coupling strength and suppresses $\tc$. With increasing pressure, the  conventional e-ph coupling $\lambda$ decreases slightly, while the vertex-corrected coupling $\lambda^{\rm V}$ increases, indicating that nonadiabatic effects become more pronounced. The FBW approach including both anharmonic phonons and vertex effects yields $\tc$ values in close agreement with experimental measurements for D$_3$S at both pressures and for H$_3$S at 200~GPa. We find that the lower phonon frequencies in D$_3$S weaken the vertex contribution and reduce $\tc$, consistent with a phonon-mediated pairing mechanism. These results show that accurate modeling of superconductivity in hydrogen-rich materials under extreme compression requires the combined treatment of anharmonic lattice dynamics, nonadiabatic corrections, and an energy-dependent electronic structure.

% Experimental section

\section{\label{sec:methods}Computational Methods}

We performed density-functional theory (DFT) calculations using Quantum {\small ESPRESSO}~\cite{Giannozzi2017} with optimized norm-conserving Vanderbilt ({\small ONCV}) pseudopotentials~\cite{Hamann2013} generated with Perdew-Burke-Ernzerhof functional~\cite{Perdew1996}. We used a plane-wave cutoff of 60~Ry, a Methfessel-Paxton smearing~\cite{Methfessel1989} value of 0.01~Ry, and a $\Gamma$-centered $k$-grid of $24 \times 24 \times 24$ to describe the electronic structure. The lattice parameters and atomic positions were relaxed until the total energy was converged within $10^{-6}$~Ry and the maximum force on each atom was less than $10^{-4}$ Ry/\AA. The dynamical matrices and the linear variation of the self-consistent potential were calculated within density-functional perturbation theory~\cite{Baroni2001} on a $4 \times 4 \times 4$ $q$-mesh. 
For the anharmonic calculations, we performed static DFT computations on a $12 \times 12 \times 12$ $k$-grid, and obtained the interatomic force constants from a $4\times 4 \times 4$ supercell with a finite displacement of 0.01~\AA. All calculations were carried out at 0~K. In our previous study we showed that the anharmonic phonon dispersion at 200~GPa remains nearly unchanged even at 200~K~\cite{Mishra2025}.

For the e-ph interactions and superconducting properties, we employed the EPW code~\cite{Lucrezi2024,Lee2023,Giustino2007,Ponce2016,Margine2013}. The electronic wavefunctions required for the Wannier interpolation~\cite{Marzari2012,Pizzi2020,Marazzo2024} were obtained on a uniform $\Gamma$-centered $8 \times 8 \times 8$ $k$-grid using ten atom-centered orbitals, with $s$ orbital for the H atom and $s, p,  d_{xy}, d_{xz}, d_{yz}$ orbitals for the S atom. The e-ph matrix elements were then computed on uniform $48 \times 48 \times 48$ $k$- and $24 \times 24 \times 24$ $q$-point grids, with an energy window of $\pm 0.2$~eV around the Fermi level. The electronic and phononic Dirac delta functions were replaced by Gaussians broadenings of 50~meV and 2.5~meV, respectively. The isotropic Eliashberg equations were solved on the imaginary Matsubara frequency axis using a frequency cutoff of 2.5~eV. For the convergence of the energy window in FBW calculations, the DOS was evaluated using a Fermi window of 1~eV. To capture the isotope effect, the same computational protocol was applied to D$_3$S by substituting the hydrogen mass with that of deuterium. 

\begin{acknowledgments}
This research was supported by the National Science Foundation (NSF) under Award No. DMR-2035518. Computational resources were provided by the Texas Advanced Computing Center (TACC) at The University of Texas at Austin through allocation TG-DMR180071 on the Stampede3 system, supported by NSF Award OAC-2311628.
\end{acknowledgments}

%\bibliography{pap}

%apsrev4-2.bst 2019-01-14 (MD) hand-edited version of apsrev4-1.bst
%Control: key (0)
%Control: author (8) initials jnrlst
%Control: editor formatted (1) identically to author
%Control: production of article title (0) allowed
%Control: page (0) single
%Control: year (1) truncated
%Control: production of eprint (0) enabled
%

\end{document}

% \title{Supplementary Information: \\Anharmonicity, electron-phonon vertex corrections, and isotope effect in high-$T_c$ superconductor H$_3$S}
\title{Supplementary Information: \\ Nonadiabatic and anharmonic effects in high-pressure H$_3$S and D$_3$S superconductors}

\author{Shashi B. Mishra}
%\email{smishra9@binghamton.edu}
\email{mshashi125@gmail.com}
\affiliation{Department of Physics, Applied Physics and Astronomy, Binghamton University-SUNY, Binghamton, New York 13902, USA}
\author{Elena R. Margine}
\email{rmargine@binghamton.edu}
\affiliation{Department of Physics, Applied Physics and Astronomy, Binghamton University-SUNY, Binghamton, New York 13902, USA}
\date{\today}

\maketitle

\begin{figure}[!hbt]
    \centering
    \includegraphics[width=0.8\textwidth]{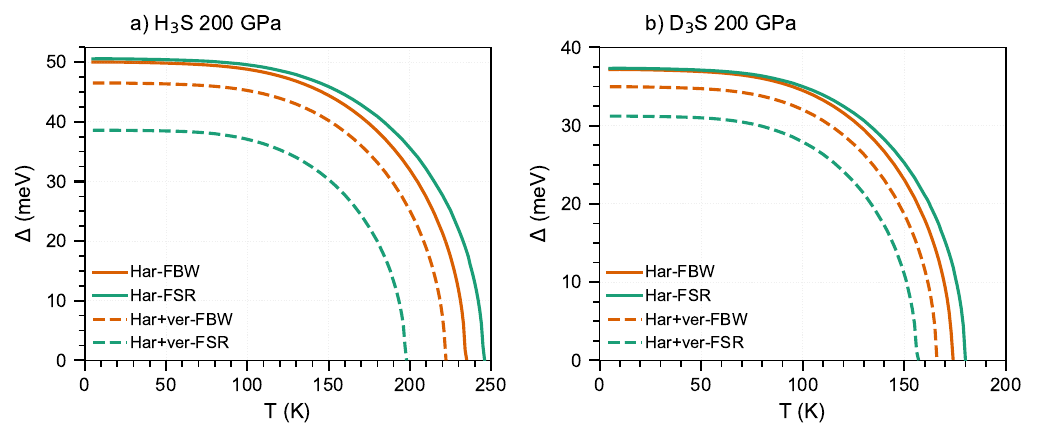}
    \caption{Isotropic superconducting gap $\Delta(T)$ calculated for a) H$_3$S and b) D$_3$S at 200~GPa using harmonic phonons and an effective Coulomb pseudopotential $\mu_{\mathrm{E}}^* = 0.16$. Results are shown for four approaches: {\small FBW} (orange), {\small FSR} (teal-green), vertex-corrected {\small FBW} (dashed orange), and vertex-corrected {\small FSR} (dashed teal-green).}
    \label{fig:gap-all}
\end{figure}

\begin{table*}[!hbt]  
    \caption{ 
    Zero-temperature superconducting gap $\Delta(0)$, critical temperature $T_{\mathrm{c}}$, and isotope coefficient $\alpha = - \ln(T_c^{\rm D_3S}-T_c^{\rm H_3S}) /{\ln(M^{\rm D_3S}-M^{\rm H_3S})}$ obtained from the FSR and FBW approaches, with and without vertex corrections for H$_3$S and D$_3$S, incorporating anharmonic phonons at 160~GPa and both harmonic and anharmonic phonons at 200~GPa. All superconductivity calculations employ a Coulomb pseudopotential $\mu_{\rm E}^*$=0.16. Experimental $\alpha$ values are taken from Ref.~\cite{Drozdov2015,Minkov2020,Du2025}. 
    }
    \label{tab:gap}
\setlength\tabcolsep{2.5pt}
\smallskip
\begin{tabular*}{\textwidth}{@{\extracolsep{\fill}}l c c c c c c c }
\hline\hline \noalign{\vskip 1mm}
Material & Pressure (GPa) & Phonons & Method & $\Delta(0)$ (meV) & $T_{\rm c}$ (K) & $\alpha$ & $\alpha^{\rm Exp}$ \\
\hline 
\multirow{12}{*}{H$_3$S}
    & \multirow{4}{*}{160} & \multirow{4}{*}{Anh} 
        & {\small FSR} & 39.69 & 205 & 0.37 & \multirow{4}{*}{0.22--0.34} \\
    & & & {\small FBW} & 44.94 & 219 & 0.38 & \\
    \cline{4-7}
    & & & ver-{\small FSR} & 36.49 & 191 &  0.34 & \\
    & & & ver-{\small FBW} & 43.49 & 213 &  0.37 & \\
    \cline{2-8}
    & \multirow{8}{*}{200} & \multirow{4}{*}{Har} 
        & {\small FSR} & 50.55 & 246 &  0.45 & \multirow{8}{*}{0.23--0.38}\\
    & & & {\small FBW} & 50.03 & 235 &  0.44 & \\
    \cline{4-7}
    & & & ver-{\small FSR} & 38.56 & 198 &  0.34 & \\
    & & & ver-{\small FBW} & 46.52 & 222 &  0.42 & \\
    \cline{3-7} 
    & & \multirow{4}{*}{Anh} 
        & {\small FSR} & 36.41 & 192 &  0.34 & \\
    & & & {\small FBW} & 37.07 & 189 &  0.33 &  \\
    \cline{4-7}
    & & & ver-{\small FSR} & 28.84 & 157 &  0.25 & \\
    & & & ver-{\small FBW} & 34.51 & 178 &  0.32 &  \\
\hline
\multirow{12}{*}{D$_3$S}
    & \multirow{4}{*}{160} & \multirow{4}{*}{Anh} 
        & {\small FSR} & 31.32 & 159 &  & \\ 
    & & & {\small FBW} & 35.05 & 168 &  & \\
    \cline{4-6}
    & & & ver-{\small FSR} & 29.53 & 151 &  & \\
    & & & ver-{\small FBW} & 34.09 & 165 &  & \\
    \cline{2-8}
    & \multirow{8}{*}{200} & \multirow{4}{*}{Har} 
        & {\small FSR} & 37.31 & 180 &  & \\
    & & & {\small FBW} & 37.17 & 174 &  & \\    
    \cline{4-6}
    & & & ver-{\small FSR} & 31.19 & 157 &  & \\
    & & & ver-{\small FBW} & 34.96 & 166 &  & \\
    \cline{3-7}
    & & \multirow{4}{*}{Anh} 
    & {\small FSR} & 29.40 & 152 &  & \\
    & & & {\small FBW} & 29.81 & 150 &  & \\
    \cline{4-6}
    & & & ver-{\small FSR} & 24.78 & 132 &  & \\
    & & & ver-{\small FBW} & 28.08 & 143 &  & \\
\hline\hline
\end{tabular*}
\end{table*}

\newpage

%\bibliography{pap}

%apsrev4-2.bst 2019-01-14 (MD) hand-edited version of apsrev4-1.bst
%Control: key (0)
%Control: author (8) initials jnrlst
%Control: editor formatted (1) identically to author
%Control: production of article title (-1) disabled
%Control: page (0) single
%Control: year (1) truncated
%Control: production of eprint (0) enabled
%